\documentclass[11pt,english,twoside]{article}
\usepackage[T1]{fontenc}
\usepackage[latin1]{inputenc}
\usepackage{graphicx}

\makeatletter

\newcommand{\noun}[1]{\textsc{#1}}

\makeatletter

\makeatletter

\renewcommand{\maketitle}{}

\makeatletter

\usepackage{asp2006}

\usepackage{lscape}

\makeatother

\makeatother

\makeatother

\usepackage{babel}
\makeatother
\begin{document}

\title{Toward a galaxian luminosity-metallicity relation spanning ten magnitudes}

\maketitle

\author{Ivo Saviane, Fabio Bresolin, John Salzer}

\affil{European Southern Observatory, Casilla 19001, Santiago 19, Chile, 
Institute for Astronomy, 2680 Woodlawn Drive, Honolulu, HI 96822
Wesleyan University, Department of Astronomy, Middletown, CT 06459} 


\begin{abstract}
We describe the two projects by which we are assembling a database
of near-IR luminosities and direct oxygen abundances for both high-
and low-mass star-forming galaxies in the nearby Universe. This will
eventually allow us to construct the first reliable and homogeneous
luminosity-metallicity relation, to be compared to theoretical models
of galactic evolution, and to the relations for galaxies at higher
redshifts.
\end{abstract}

\section{Introduction}


The mass-metallicity (M-Z) relation is a powerful diagnostic of galaxy
evolution, as first shown by Larson (\citeyear{larson74}). In this pioneering
paper, the relation for elliptical galaxies was reproduced via galactic
winds powered by SNe explosions and feedback, and cooling was provided
by SN remnants. Recently, the first attempts to predict the M-Z relation
in the framework of hierarchical models of structure formation have
been seen as well (e.g., Tissera et al. \citeyear{tissera_etal05}; T05).
The time-dependence of the relation has also been explored by T05,
and their simulations show that a well-defined relation should be
present as early as $z=3$. This redshift corresponds to $\sim1.5$~Gyr
after the big bang, when only $\sim10\%$ of the stellar mass in the
Universe had been formed, according to Dickinson et al. (\citeyear{dickinson_etal03}).
This would mean that the M-Z relation is established very early in
the life of galaxies. An important test of the theory is then to check
that the predicted M-Z relation is consistent with observations both
in the local and in the high-redshift Universe. Since luminosities
are easier to measure than masses for large samples of galaxies, the
luminosity-metallicity (L-Z) relation is often utilized instead of
the one between mass and metallicity. Indeed in this contribution
we discuss our project to obtain the best possible L-Z relation for
the local Universe.

\subsection{Which abundances to use?}

When defining an empirical L-Z relation, one can measure either the
abundance of the gas or that of the stars contained in each galaxy.
However, when testing the \emph{present} relation, gas abundances
provide less ambigous results. This is because interstellar medium
(ISM) abundances can be most easily obtained from the emission-line
spectra of the gas ionized by massive young-hot stars, and these metallicities
are those of the most recent stellar generation. Therefore they can
be compared to the current \emph{}galaxian luminosity. Moreover, in
nearby galaxies (both giant and dwarf) individual \ion{H}{ii} regions
can be measured, and one can study the radial dependence of abundances.
Finally, only moderate spectroscopic resolution is needed to measure
emission-line ratios, so ISM abundances can be obtained out to relatively
high redshifts, thanks also to the large fluxes involved (see Fig.~\ref{fig:The-flux-calibrated-and}).

On the contrary, a non-ambiguous stellar L-Z can be obtained only
if we can estimate the age of the stars that we are measuring. This
can be done only for nearby resolved galaxies, where stellar ages
can be determined (with variable degrees of accuracy) by comparing
their position in the color-magnitude diagram with the predictions
of theoretical isochrones (e.g., Pont et al. \citeyear{pont_etal04}).
Moreover, the need for high resolution spectroscopy limits this method
to the dwarf satellites of the Milky Way. Low-resolution spectroscopy
calibrated to high-resolution abundances, or the use of stellar colors,
help extending this method to galaxies of the Local Group, but the
metallicities obtained in this way have large uncertainties. In all
other cases only integrated spectroscopy or photometry can be used,
and since the SEDs of distant galaxies are the convolution of those
of their various stellar generations, the spectroscopic parameters
that we measure will yield an age-weighted metallicity. A meaningful
comparison with a theoretical L-Z relation thus cannot be done, because
the SFH is not known. 



\subsection{Which luminosities to use?}


As it was mentioned above, luminosities are used as proxies for the
masses, and to estimate the current mass of a galaxy we must have
an idea of its $M/L$ ratio. For simple stellar populations the $M/L$
ratio increases almost linearly with age (e.g., Bruzual \& Charlot
\citeyear{bc03}; BC03), and for a complex stellar population the ratio
will have a value somewhere between the minimum and maximum. It is
then important to measure luminosities within a photometric band that
minimizes the variations of $M/L$ with age. In this respect near-IR
(NIR) luminosities have a clear advantage over optical luminosities,
since, for example, $M/L_{V}$ is four times more sensitive to the
age of the population than $M/L_{K}$ (BC03). A NIR luminosity will
then smooth out the galaxy-to-galaxy differences in SFH better than
an optical luminosity. Moreover, the effects of internal absorption
in measuring the luminosity are greatly reduced as well (Salzer et
al. \citeyear{salzer_etal05}).

\section{Toward a universal L-Z relation }

\begin{figure}[t]
\begin{centering}\includegraphics[width=0.85\textwidth]{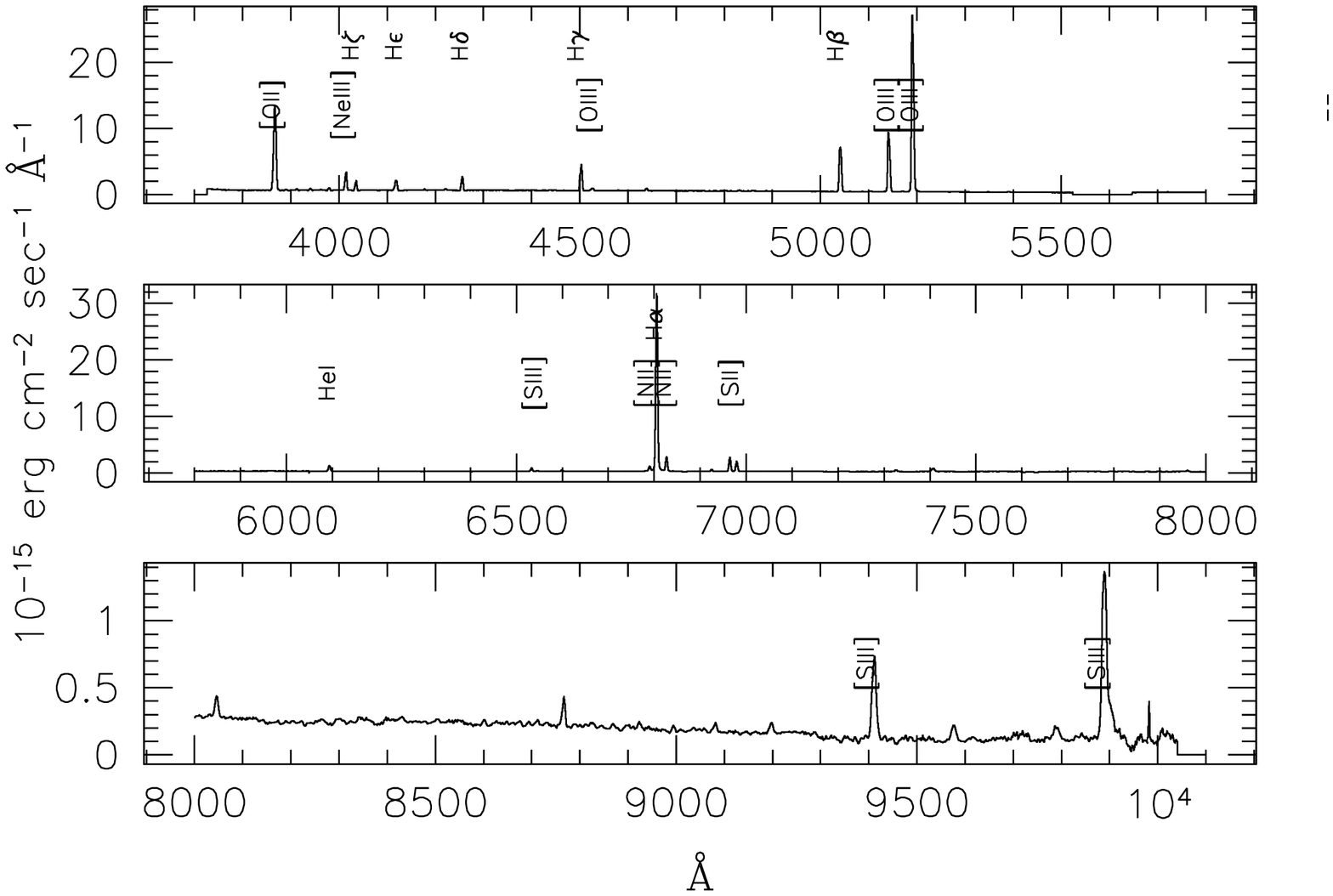}\par\end{centering}

\caption{The flux-calibrated and merged spectrum of KISS \#242. The main emission
lines are identified. \label{fig:The-flux-calibrated-and}}
\end{figure}

It should now be clear that the less ambiguous L-Z relation in the
local Universe should be based on homogeneous ISM abundances and NIR
luminosities. However until recently it was not possible to assemble
such a relation based on literature data. On the one side reliable
abundances for the high-mass galaxies were lacking, and on the other
side NIR luminosities for the low-mass galaxies were hard to find.
Our project was then started to try and remedy this situation. 

With respect to the abundances, one must remark that oxygen is the
most abundant metal, and that its emission lines are the most prominent
metal lines in the optical spectrum of an \ion{H}{ii} region (see
Fig.~\ref{fig:The-flux-calibrated-and}). Oxygen is then the element
that is usually discussed. Abundance determination methods can be
quite sensitive to the electron temperature, so a reliable measurement
requires determination of $T_{{\rm e}}$. Traditionally this is computed
using the ratio of oxygen lines {[}\ion{O}{iii}]$\lambda4363/${[}\ion{O}{iii}]$\lambda\lambda4959,5007$,
however when the oxygen line fluxes start to drop at high metallicities
(e.g., McGaugh \citeyear{mcgaugh91}), the weak {[}\ion{O}{iii}]$\lambda4363$
line becomes more and more difficult to measure. It is then commonplace
to switch from this so-called `direct' method to the `empirical' method,
which consists of measuring only the strong lines [\ion{O}{ii}]$\lambda3727$
and [\ion{O}{iii}]$\lambda\lambda4949,5007$, and finding the oxygen
abundance via a calibration obtained with \ion{H}{ii} regions of
known metallicity and photoionization models (e.g., Edmunds \& Pagel
\citeyear{edmunds_pagel84}). Aside from the lower accuracy, the problem
with these abundances is that different calibrations exist for the
empirical method (e.g., Salzer et al. \citeyear{salzer_etal05}), yielding
very different final values. Since all L-Z relations for giant galaxies
use empirical oxygen abundances, we started a project to find direct
abundances for a sample of star-forming galaxies from the KISS survey,
exploiting temperature-sensitive line ratios for elements other than
oxygen. The KISS project (e.g., Salzer et al. \citeyear{salzer_etal00})
has collected NIR luminosities for all the galaxies in the catalog
that were detected in the 2MASS survey, so once robust abundances
become available, we will be able to make the first reliable comparisons
with theoretical L-Z relations.

The second critical point mentioned above is the lack of NIR luminosities
for low-mass galaxies. This problem is being tackled in a parallel
effort, where we have been collecting homogeneous, direct oxygen abundances
and NIR luminosities for a sample of nearby dwarf star-forming galaxies.
Dwarf galaxies have low surface brightnesses, so NIR observations
from the ground require long observational programs. For this reason,
until recently catalogs lacked information on dwarf galaxy infrared
luminosities. Moreover, while for distant galaxies the redshift can
be used to estimate their distances, peculiar velocities in the nearby
Universe have a non-negligible effect, and they can greatly affect
the accuracy of luminosity measurements. Therefore we concentrated
our efforts on dwarf irregular (dIrr) galaxies in nearby groups, where
distances are the same, and the existence of an L-Z relation can be
checked even using apparent magnitudes. Reports of this work in progress
have been given elsewhere (e.g., Saviane et al. \citeyear{saviane_etal05}),
and the first paper of the series has been submitted. In the next
section we then limit our discussion to the project dealing with direct
abundances of KISS galaxies.




\subsection{Direct abundances of star-forming galaxies}

\begin{figure}
\begin{centering}\includegraphics[width=0.6\textwidth]{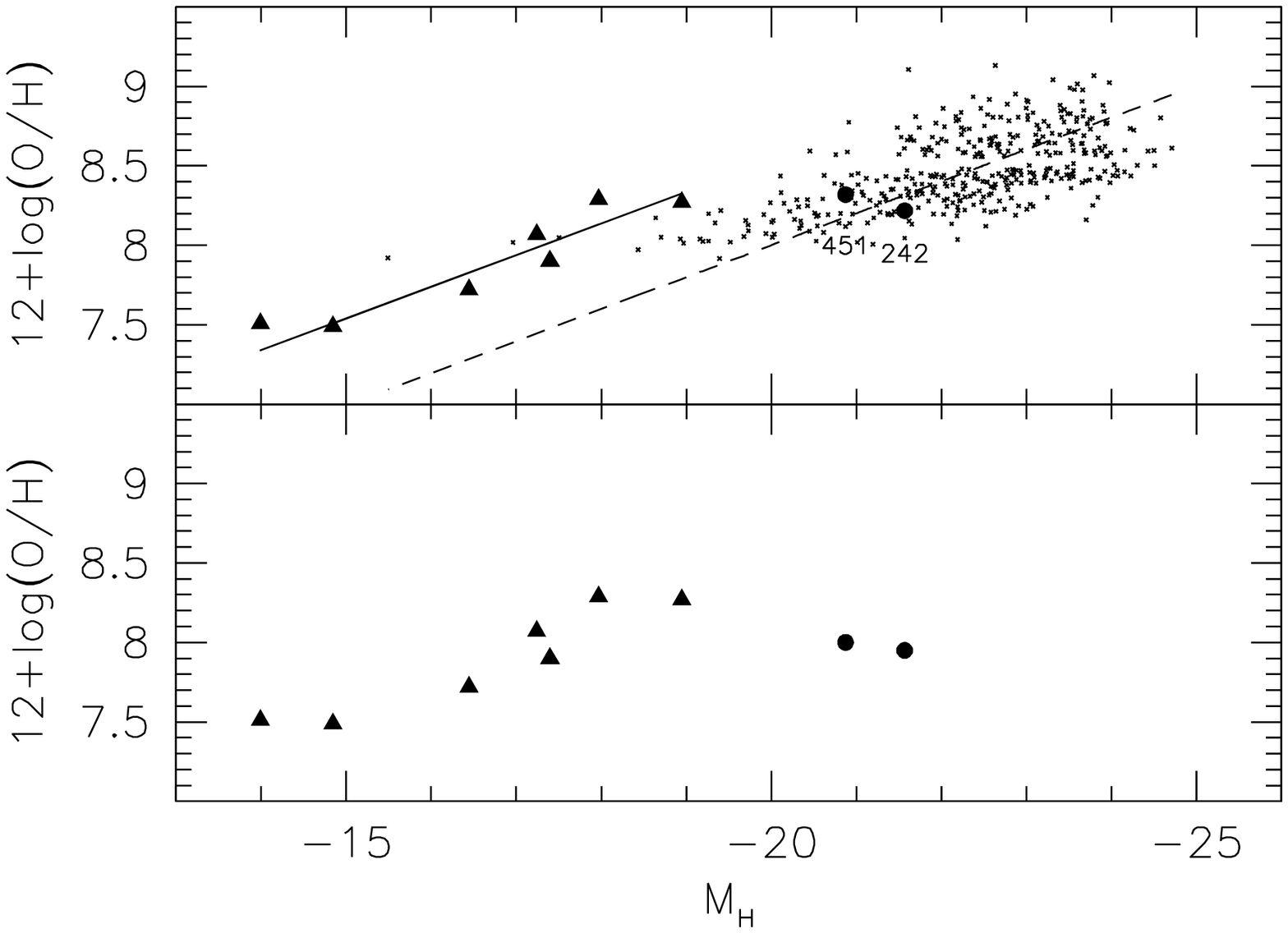}\par\end{centering}

\caption{The small crosses in the top panel show the location of KISS galaxies
in the L-Z plane, when empirical abundances are used. The dashed line
is the fit obtained by Salzer et al. (\citeyear{salzer_etal05}). Filled
circles and labels identify galaxies discussed in the text. The filled
triangles and solid line represent the L-Z relation obtained with
dIrr galaxies in nearby groups. In the lower panel the  filled circles
represent the same KISS galaxies of the top panel, when direct abundances
are used. \label{fig:The-open-circles}}
\end{figure}

Aside from the oxygen line ratio quoted above, the electron temperature
can be computed via additional line ratios that can be formed with
optical emission lines, e.g., the {[}\ion{S}{iii}] ratio $\lambda6312/\lambda\lambda(9069,9532)$
and the [\ion{N}{ii}] ratio $\lambda5755/\lambda\lambda(6584,6583)$.
These other ion lines are measurable in metal-rich galaxies with 10m-class
telescopes (e.g., Bresolin \citeyear{bresolin06}), and in particular
{[}\ion{N}{ii}] lines stay in the optical domain up to redshifts
$z\simeq0.5$. Our project is then to measure reliable abundances
even in the more massive galaxies, using these line ratios. To do
this, a relatively high resolution ($\sim1$\AA/px) is needed to
detect the faint lines and deblend e.g., the {[}\ion{N}{ii}] lines
from H$\alpha$. At the same time a complete coverage of the optical
domain (from $0.37$~$\mu$m to $1$~$\mu$m) must be ensured. These
requirements, together with the northern location of the KISS galaxies,
made us choose LRIS at the W.~M.~Keck telescope to acquire our spectra.
The initial target list was selected by imposing a range in the empirical
abundances $8.2<12+\log({\rm O/H})<8.9$ and a range of luminosities
$-20.8<M_{H}<-25$. The spectral coverage was ensured by taking blue
spectra with grism 600/4000, and red spectra with gratings 400/8500
and 900/5500. The spectra were reduced with standard procedures within
the the \noun{eso/midas} environment. An example of an extracted,
merged, and flux-calibrated spectrum is given in Fig.~\ref{fig:The-flux-calibrated-and}.
Line fluxes were measured by simple integration within the profile,
and they were corrected for internal reddening by comparing the H$\alpha/$H$\beta$
ratio to the theoretical value from Hummer \& Storey (\citeyear{hummer_storey87}).
Chemical abundances were computed using the \noun{stsdas.nebular}
package in \noun{iraf} (Shaw \& Dufour \citeyear{shaw_dufour95}). Empirical
abundaces were estimated as well, using the Kennicutt et al. (\citeyear{kbg03})
calibration.

Of the fifteen galaxies that we observed, we discuss here the two
galaxies with available $H$ magnitudes and where {[}\ion{O}{iii}]$\lambda4363$
could be measured, together with the other lines entering the temperature-sensitive
ratios discussed above. In Fig.~\ref{fig:The-open-circles} the central
abundances vs. $H$ absolute magnitudes are plotted for these galaxies
and for the nearby dwarfs. When we compare the abundances for the
two KISS galaxies measured with the direct method with their corresponding
empirical estimates based on their strong lines, we find that the
latter values over-estimate the actual abundances in both cases. Considering
only direct abundances (lower panel), the L-Z relation seems to flatten,
but more galaxies need to be included to confirm this trend, especially
in the high luminosity range. Indeed selecting galaxies with a measurable
{[}\ion{O}{iii}]$\lambda4363$ tends to select objects at the lowest
${\rm O/H}$ range. A similar conclusion is reached if we compare
our L-Z relation to that of the SDSS survey (Tremonti et al. \citeyear{tremonti_etal04}).
Most galaxies in that relation have super-solar metallicities, which
are possibly overestimated due to the empirical method that was used,
as suggested by Bresolin (\citeyear{bresolin06}).


\section{Discussion}

Until recently, L-Z relations for nearby star-forming galaxies were
based on optical luminosities and empirical oxygen abundances, which,
for the reasons discussed above, are not suitable to define a reliable
trend of metallicity with luminosity. Fortunately in the last few
years better data sets have started appearing in the literature. In
particular Salzer et al. (\citeyear{salzer_etal05}) have published relations
for massive star-forming galaxies based on NIR imaging (2MASS) and
empirical abundances computed with different calibrations; and Vaduvescu
et al. (\citeyear{vaduvescu_etal05,vaduvescu_etal06}) have obtained NIR
luminosities for dwarf star-forming galaxies in the local volume and
in the Virgo cluster. Moreover Lee et al. (\citeyear{lee_etal06}) have
published a NIR L-Z relation for a number of nearby dIrr galaxies
having direct oxygen abundances. Since their luminosities are in the
Spitzer $4.5~\mu{\rm m}$ band, their L-Z relation cannot be compared
to Salzer et al. directly. Instead they convert it into a mass-luminosity
relation and try and extend the SDSS relation to low-mass galaxies.
Given our previous discussion, it would seem like this exercise has
a somewhat limited value, since it is mixing direct and empirical
abundances in the same relation. We believe that it is more appropriate
to try and get direct oxygen abundances in a significant sample of
star-forming galaxies. This is indeed the aim of the project presented
in this contribution, which complements our parallel effort to obtain
direct abundances and NIR luminosities of dwarf star-forming galaxies.
A reliable local calibration of the L-Z relation will be fundamental
to properly interpret the higher redshift relations that are being
published (Kobulnicky \& Kewley~\citeyear{kobulnicky_kewley04}, Tremonti
et al.~\citeyear{tremonti_etal04}, Maier et al.~\citeyear{maier_etal05},
Savaglio et al.~\citeyear{savaglio_etal05}, Erb et al.~\citeyear{erb_etal06}).
In particular we have no good handle on the amount of chemical evolution
that is being implied by the high-z studies, since their point of
overlap with the low-z samples is only at high metallicities, where
the calibration is so uncertain. Another key motivation for our study
is then to provide a direct calibration of the strong line method
at higher abundances. This is vital to place all data sets, both at
high and low redshift, on a uniform and reliable metallicity scale. 

Though at this stage it is premature to draw definite conclusions,
our data are showing that abundances for high-mass galaxies have probably
been overestimated in the past. To check this result, direct abundances
for all the galaxies in our sample will be computed using the {[}\ion{S}{iii}]
and the [\ion{N}{ii}] line ratios. NIR imaging will also be acquired
for the targets not included in 2MASS. 


\acknowledgements 

We thank Sandra Savaglio for a careful reading of an early draft of
this manuscript.

\end{document}